\documentclass{article}
\usepackage{graphicx} 
\usepackage[
colorlinks=false
]{hyperref}
\usepackage[
backend=biber,
style=alphabetic,
giveninits=true,
]{biblatex}
\usepackage{authblk}
\usepackage{amsmath,amssymb}
\usepackage{hyperref}
\usepackage[colorinlistoftodos]{todonotes}
\usepackage{mathtools}

\AtEveryBibitem{
    \clearfield{urlyear}
    \clearfield{urlmonth}
    \clearfield{month}
    \clearfield{day}
}

\title{A simple, accurate model for detachment access}

\author[1]{Thomas Body}
\author[2]{Arne Kallenbach} 
\author[1]{Thomas Eich}
\affil[1]{Commonwealth Fusion Systems}
\affil[2]{Max Planck Institute for Plasma Physics, Garching}

\date{\today}

\begin{document}
\maketitle

\section{Abstract}\label{sec:abstract}

In next-step fusion tokamaks such as SPARC and ITER, achieving high levels of scrape-off-layer power dissipation will be essential to protect the divertor while maintaining good core plasma performance. The Lengyel model for power dissipation is easy to interpret and fast enough to incorporate into plasma control and scoping tools, but it systematically overestimates the impurity concentration required to reach detachment by a factor of $\sim5$ relative to experiments and higher-fidelity simulations. In this work, we extended the Lengyel model to match the semi-empirical Kallenbach scaling, which successfully describes detachment access on several operating tokamaks. We found that we can reproduce the experimental scaling by accounting for cross-field transport in the divertor, power and momentum loss due to neutral ionization close to the divertor target and turbulent broadening of the upstream heat flux channel. These corrections cause the impurity concentration required for detachment to decrease faster than $n_{e,u}^2$, reproducing the $c_z\propto 1/n_{e,u}^{2.7-3.2}$ scalings found in experiment. The model also quantitatively reproduces the impurity concentration needed to reach detachment in experiment, demonstrating that the extended Lengyel model can be used as a simple, accurate model for detachment access.

\section{Introduction}\label{sec:introduction}

To produce net energy from fusion, we need to keep the center of the plasma at temperatures about ten times hotter than the center of the sun, while simultaneously protecting the solid device walls from melting and cracking. These two requirements place opposing demands on tokamak operation, and the challenge of core-edge integration is to find an optimal middle ground. While already challenging for existing devices, finding scenarios which maximize fusion performance while maintaining tolerable heat exhaust will be even more critical as we push beyond fusion breakeven in next-step tokamaks such as SPARC and ITER. Relative to existing devices, these tokamaks are predicted to need more power crossing the separatrix to stay in H-mode \cite{Martin2008-vw}, and this power must be exhausted from a narrower heat flux channel \cite{Eich2013-yg}. Combining these two effects, the upstream parallel heat flux density is predicted to scale as $q_\parallel\sim\frac{P_{sep}}{A_{wetted}}\sim B_T^{2.52}R^{0.16}$ for constant $f_{LH}=P_{sep}/P_{LH}$, constant shaping and constant $q_{95}$  \cite{Reinke2017-qx}. The heat flux entering the scrape-off-layer must either be absorbed at the divertor targets, or radiated as a photon flux and absorbed over the first wall. The heat flux reaching the divertor targets must be kept below material limits, beyond which top surface melting and tile cracking can occur. The sheath-entrance plasma temperature must also be reduced, since high sheath-entrance plasma temperatures will cause excessive tungsten sputtering from the divertor targets.

To mitigate the heat flux to the divertor in next-step tokamaks, large fractions of the power entering the scrape-off-layer will need to be radiated. To increase the fraction of power radiated, impurities such as neon or argon will be injected into the scrape-off-layer. Even at concentrations of a few percent, these impurities radiate much more power than the hydrogen main ions because they have bound electrons at higher temperatures \cite{Putterich2019-uq}. With sufficient seeding, the plasma recombines into neutral gas before it reaches the divertor targets. This `detachment' greatly reduces the heat flux to the divertor targets \cite{Krasheninnikov2017-ay}, protecting the divertor from melting or cracking and greatly reducing tungsten sputtering.

Impurity seeding must be balanced against its impact on the core plasma. A fraction of the impurities seeded into the divertor will reach the confined region where they radiatively cool the core plasma and dilute the fuel ions, causing a steep drop in the fusion rate. Finding core-edge integrated scenarios relies on two key aspects: protecting the divertor while minimizing impurity seeding, and maximizing the ratio of the divertor impurity concentration to the core impurity concentration \cite{Kallenbach2024-as}. Impurity enrichment is challenging to predict for future devices, and as such, we focus on developing models to predict the impurity concentration required for power dissipation.

Sophisticated scrape-off-layer `transport' models such as SOLPS-ITER have been developed \cite{Wiesen2015-hc}, validated \cite{Wensing2021-tj,Horsten2025-om}, and used to predict the impurity concentration required for power dissipation in SPARC \cite{Ballinger2021-aq,Lore2024-dv} and ITER \cite{Lore2022-qj,Veselova2021-yf}. However, performing simulations with (or training neural networks of \cite{Dasbach2023-fk,Wiesen2024-ns}) transport models remains computationally expensive and labor intensive, and interpreting results from complex multi-physics models is not straightforward. In this paper, we go in the opposite direction and ask what is the simplest model we can use for power exhaust that remains reasonably accurate?

A promising candidate is the `Lengyel model' \cite{Lengyel1981-in} (derived in section \ref{sec:extended_lengyel}). This model predicts the impurity fraction $c_z$ required to radiate some desired fraction $f_{pow,SOL}\approx f_{rad,SOL}$ of the parallel heat flux density $q_\parallel$ directed towards one of the divertor targets. In this work, we focus on outer divertor detachment, since the inner divertor usually detaches before the outer divertor \cite{Krasheninnikov2017-ay} in favorable drift direction. The fixed-fraction scrape-off-layer impurity concentration (the impurity ion density divided by the electron density) required for power dissipation is
\begin{align}
c_z =\frac{n_{z}}{n_e} &= \frac{q_{\parallel,u}^2 - q_{\parallel,t}^2}{2 \kappa_e n_{e,u}^2T_{e,u}^2 L_{INT}}\label{eq:basic_lengyel}
\end{align}
where
\begin{align}
q_{\parallel,t} &= (1-f_{pow,SOL}) q_{\parallel,u}\\
L_{INT}&= \int_t^u L_z(T_e) \sqrt{T_e} \partial T_e
\end{align}
for a given upstream parallel heat flux density $q_{\parallel,u}$ (defined in equation \ref{eq:qpar_definition}), parallel electron heat conductivity $\kappa_e$ (defined in equation \ref{eq:kappa_e}), upstream electron density $n_{e,u}$ and upstream electron temperature $T_{e,u}$. The $L_{INT}$ term is an integral along a flux-tube with respect to the electron temperature $T_e$, from the sheath-entrance electron temperature $T_{e,t}$ to the upstream electron temperature $T_{e,u}$. The $L_z$ term is a temperature-dependent cooling factor \cite{Putterich2019-uq} defined such that the impurity radiated power per unit volume will be
\begin{align}
    P_{rad}/V &= c_z n_e^2 L_z(T_e)\label{eq:Lz_definition}
\end{align}
These cooling factors are computed from OpenADAS ADF-11 data \cite{Summers2006-rp} for $n_e\tau=0.5 \mathrm{ms}\times10^{20}\mathrm{m^{-3}}$ and $n_e=10^{20}\mathrm{m^{-3}}$ using \href{https://github.com/cfs-energy/radas/}{radas}\footnote{Available at \href{https://github.com/cfs-energy/radas/}{\texttt{github.com/cfs-energy/radas/}}.}. Despite its simplicity, the Lengyel model gives similar results to 1D Braginskii simulations using fixed-fraction impurities \cite{Body2024-ky}. However, comparisons to 2D Braginskii simulations with impurity transport show that the Lengyel model needs significantly higher impurity concentrations to reach detachment than what is found in the higher-fidelity simulations \cite{Moulton2021-id,Jarvinen2023-xh}. Surprisingly, the ratio between the Lengyel and SOLPS results was a constant factor of $4.3$, and using the Lengyel model with this calibration factor reproduced the SOLPS results with remarkable accuracy \cite{Moulton2021-id}. Comparisons to experiment found similar calibration factors, between 5.5 on ASDEX Upgrade and 2.9 on JET \cite{Henderson2021-ts}. This leads to an intriguing possibility: if we could self-consistently predict these calibration factors, the Lengyel model would provide an accurate, fast way of computing the impurity concentration required for detachment.

Another simple model for power dissipation is the semi-empirical `Kallenbach scaling' \cite{Kallenbach2015-uu,Kallenbach2016-hk}. This scaling predicts the degree of detachment $q_{det}$ for mixed impurity seeding \cite{Henderson2023-hu}
\begin{align}
    q_{det} &= \frac{1.3}{1 + \sum_z f_z c_z}\frac{P_{sep} / \mathrm{MW}}{R_0 / \mathrm{m}}\frac{\mathrm{Pa}}{p_{div}}\frac{5\mathrm{mm}}{\lambda_{INT}}
    \label{eq:kallenbach_scaling}
\end{align}
where
\begin{align}
    q_{det}&=
\begin{cases}
<1 \text{ indicates pronounced detachment,}\\
\sim1\text{ indicates partial detachment,}\\
>1\text{ indicates attachment.}
\end{cases}
\end{align}
and
\begin{align}
f_Z &=
\begin{cases}
18 \text{ for nitrogen,}\\
45 \text{ for neon,}\\
90 \text{ for argon.}
\end{cases} \label{eq:fZ_kallenbach}
\end{align}
for a given power crossing the separatrix $P_{sep}$, major radius $R_0$, divertor neutral pressure $p_{div}$ and integral heat flux width $\lambda_{INT}\approx \lambda_q + 1.64S$ \cite{Makowski2012-gx}. Although originally derived for nitrogen-seeded ASDEX Upgrade pulses, the Kallenbach scaling also describes detachment onset on MAST-U \cite{Henderson2024-bf}, on JET \cite{Henderson2021-ts,Henderson2024-bw} and in SOLEDGE3X simulations of TCV \cite{Yang2023-mr}. The scaling also predicts the detachment onset in mixed argon, neon and nitrogen seeding experiments on ASDEX Upgrade — albeit with slightly different $f_Z$ values ($f_{Ne}\approx32.2, f_{Ar}\approx187.6$) than those given in equation \ref{eq:fZ_kallenbach}  \cite{Henderson2023-hu}.
This demonstrates that the Kallenbach scaling accurately predicts the impurity concentration needed to reach detachment across a range of conditions, but since the scaling is semi-empirical, it doesn't directly tell us why that is the case. This provides something of a technical motivation for this paper: without understanding the mechanisms leading to detachment onset, we have less confidence that the scaling holds outside of the region where it has been validated.

The true motivation for this paper, however, is to resolve two key puzzles. The first of these is: what is the origin of the Lengyel calibration factor? The second is how does the detachment onset scale with density? If we assume $q_{\parallel,t}\approx0$, $L_{INT}\propto T_{e,u}$ (from Reinke et al., 2017 \cite{Reinke2017-qx}) and $T_{e,u}\sim q_{\parallel,u}^{2/7}$ we can simplify the Lengyel model to
\begin{align}
    c_z \propto \frac{q_{\parallel,u}^{8/7}}{{n_{e,u}^2}}\label{eq:lengyel_density_dependence}
\end{align}
If we crudely estimate $P_{sep}/(R_0 \lambda_{INT})\propto q_{\parallel,u}$ and $p_{div}\propto n_{e,u}^{1/0.31}$ (using the empirical scaling from Kallenbach et al., 2018 \cite{Kallenbach2018-ct}) we can simplify the Kallenbach scaling (for $q_{det}=1$, in the limit that $f_z c_z >>1$) to
\begin{align}
    c_z \propto \frac{q_{\parallel,u}}{n_{e,u}^{3.22}}\label{eq:kallenbach_density_dependence}
\end{align}
Under these approximations the models have similar forms, so why is the Kallenbach scaling significantly more accurate than the Lengyel model? Does the stronger-than-$n_{e,u}^2$ decrease in the impurity concentration extrapolate to higher density devices such as SPARC?

This paper addresses these questions by extending the Lengyel model to reproduce the Kallenbach scaling. In section \ref{sec:kallenbach_reformulated}, we present the semi-analytical Kallenbach model which we use as a reference for extending the Lengyel model. In section \ref{sec:solving_the_kallenbach_model}, we show that the Kallenbach and Lengyel models agree throughout most of the scrape-off-layer, but they diverge close to the divertor targets once convective transport becomes significant. In section \ref{sec:convective_scalings}, we compute numerical fits for the power and momentum loss in the convective transport region, to account for these losses in the Lengyel model. In section \ref{sec:extended_lengyel}, we use these fitted loss functions together with an approximation for divertor broadening to derive an extended Lengyel model which agrees with the Kallenbach model. In section \ref{sec:comparing_to_experiment} we compare our extended model against three experimental results — the experimental Kallenbach scaling, the detachment onset scalings from Henderson et al., 2021 \cite{Henderson2021-ts} and the values from an experimental data point from Kallenbach et al., 2024 \cite{Kallenbach2024-as} — demonstrating that our extended model is a useful tool for interpreting experimental data. Finally, in section \ref{sec:conclusion}, we discuss how the model could be further validated and extended to improve its accuracy, and provide links to a open-source implementation of the model in section \ref{sec:software_availability}.
   
\section{The Kallenbach model}\label{sec:kallenbach_reformulated}

To extend the Lengyel model to reproduce the Kallenbach scaling, we use as a reference the time-independent semi-analytical 1D model from Kallenbach et al., 2016 \cite{Kallenbach2016-hk}. This model, which we'll call the `Kallenbach model', reproduces the experimental `Kallenbach scaling' closely. As such, if we can extend the Lengyel model to match the Kallenbach model, we expect the extended model to also match the experimental scaling. Although the focus of this paper is on the Lengyel model, we'll introduce the Kallenbach model here to present the model equations in a form which can be directly compared to the Lengyel equations.

The Kallenbach model is based on equations 9.84, 9.85 and 9.86 from Stangeby, 2000 \cite{Stangeby2000-ht}, giving coupled differential equations for the conservation of particles, momentum and energy. The model includes a simple model for the neutral recycling flux from the target, which provides a particle source and an energy and momentum sink close to the target. The model assumes that a radiating impurity such as nitrogen is present with a concentration fixed to some fraction of the electron density, and this is used to calculate the radiative dissipation of the parallel heat flux. To account for broadening of the heat flux in the divertor due to cross-field transport (i.e. the power spreading factor $S$ from reference \cite{Eich2013-yg}), the flux tube radial width is assumed to step-change from $\lambda_q$ above the X-point to $\lambda_{INT}=\lambda_q+1.64S$ \cite{Makowski2012-gx} in the divertor, which leads to a drop in the parallel heat flux density $q_\parallel$. For simplicity, equal ion and electron temperatures are assumed, and ion viscosity and SOL currents are neglected.
The model consists of 5 coupled differential equations which give the spatial evolution of plasma and neutral parallel profiles as we move along the magnetic fieldline from the divertor target to the outboard midplane. For the plasma density $n$, the plasma bulk flow velocity $v$, the neutral density $n_n$, the total parallel heat flux density $q_{tot}$ and the electron temperature $T_e$, the change with respect to the parallel coordinate $x$ is;
\begin{align}
\frac{d n_e}{dx} &= \frac{1}{m_i v_i^2 - 2T_e}\left( m_i v_i (2 R_{iz} - R_{rec} + R_{cx}) + 2n_e\frac{dT_e}{dx} \right)\label{eq:kallenbach_dndx}\\
\frac{dv_i}{dx} &= \frac{1}{n_e}\left(-v_i \frac{dn_e}{dx} + R_{iz} - R_{rec}\right)\label{eq:kallenbach_dvdx}\\
\frac{dn_n}{dx} &= \frac{1}{v_n}\left(-R_{iz} + R_{rec}\right)\label{eq:kallenbach_dnndx}\\
\frac{d q_{tot}}{dx} &= n_e^2 c_z L_z(T_e) + T_i R_{cx} + E_{iz}R_{iz}\label{eq:kallenbach_dqtotdx}\\
\frac{dT_e}{dx} &= \frac{q_{cond}}{\kappa_{e} T_e^{5/2}}\label{eq:kallenbach_dTedx}
\end{align}
with
\begin{align}
v_n &= \frac{1}{4}\sqrt{\frac{8}{\pi} \frac{E_{FC}}{m_i}}\\
q_{cond} &= q_{tot} - q_{conv}\\
q_{conv} &= (5T_e + \frac{1}{2}m_iv^2)nv\\
\kappa_e &= 2390 \mathrm{W m^{-1} eV^{-3.5}}/\kappa_z\label{eq:kappa_e}
\end{align}
using the $Z_{eff}$ correction to the heat conductivity from Brown and Goldston, 2021 \cite{Brown2021-if}
\begin{align}
\kappa_z &= 0.672 + 0.076 \sqrt{Z_{eff}} + 0.252 Z_{eff}\label{eq:kappa_z}\\
Z_{eff}&=1 + c_z\langle Z \rangle \left(\langle Z \rangle -1\right)
\end{align}
where $m_i$ is the ion mass, $R_{iz}$ is the ionization rate, $R_{rec}$ is the recombination rate, $R_{cx}$ is the charge exchange rate, $v_n$ is the neutral bulk flow velocity (assumed to be mean velocity at the Franck-Condon energy $E_{FC}=5\mathrm{eV}$), $L_z$ is the impurity cooling factor introduced in equation \ref{eq:Lz_definition}, $T_i$ is the ion temperature (assumed to be equal to the electron temperature) and $E_{iz}=15.0\mathrm{eV}$ is the effective ionization energy of hydrogen. These equations are combined with sheath boundary conditions;
\begin{align}
    v_{i,t} &= (1 - \epsilon_t)c_{s,t}\label{eq:kallenbach_vt}\\
    n_{e,t} &= \frac{q_{tot,t}}{\gamma_{sh} T_{e,t}c_{s,t}}\label{eq:kallenbach_nt}\\
    n_{n,t} &= \frac{n_{e,t} c_{s,t}}{v_n}\label{eq:kallenbach_nnt}
\end{align}
where $\gamma_{sh}\approx8$ is the sheath heat transmission coefficient, $c_{s,t} = \sqrt{\frac{2T_{e,t}}{m_i}}$ is the sound speed at the target and $\epsilon_t=10^{-6}$ is a small value to prevent a division-by-zero error in the continuity equation (equation \ref{eq:kallenbach_dndx}). The set of equations are solved as an initial value problem. The total sheath heat flux $q_{tot,t}$, the sheath-entrance electron temperature $T_{e,t}$ and the impurity fraction $c_z$ are given as inputs. Equations \ref{eq:kallenbach_vt}-\ref{eq:kallenbach_nnt} are used to find a consistent set of boundary conditions at the divertor target, and equations \ref{eq:kallenbach_dndx}-\ref{eq:kallenbach_dTedx} are integrated from the divertor to the midplane, giving 1D solutions along a magnetic fieldline\footnote{The model presented above has small modifications to the model presented in Kallenbach et al., 2016 \cite{Kallenbach2016-hk}. We reformulated the continuity equation (equation \ref{eq:kallenbach_dndx}) to avoid explicitly tracking the change in the convective heat flux, which let us use an implicit multi-step ODE method suitable for stiff problems. Using a backwards differentiation formula (BDF) method reduced the computational cost by about a factor-of-ten relative to the explicit Euler integration used in Kallenbach et al., 2016 \cite{Kallenbach2016-hk}, to about $25\mathrm{ms}$ per evaluation. We used an updated $Z_{eff}$ correction (equation \ref{eq:kappa_z}), instead of the original $\kappa_z = Z_{eff}^{0.3}$, although this didn't have a significant effect on the results. We also disabled the option to increase the neutral velocity above the Franck-Condon velocity. We verified that, by disabling this term in the original implementation, our reformulated model and the original implementation match exactly.}.

\section{Comparing the Kallenbach and Lengyel models}\label{sec:solving_the_kallenbach_model}

\begin{figure}
    \centering
    \includegraphics[width=0.8\linewidth]{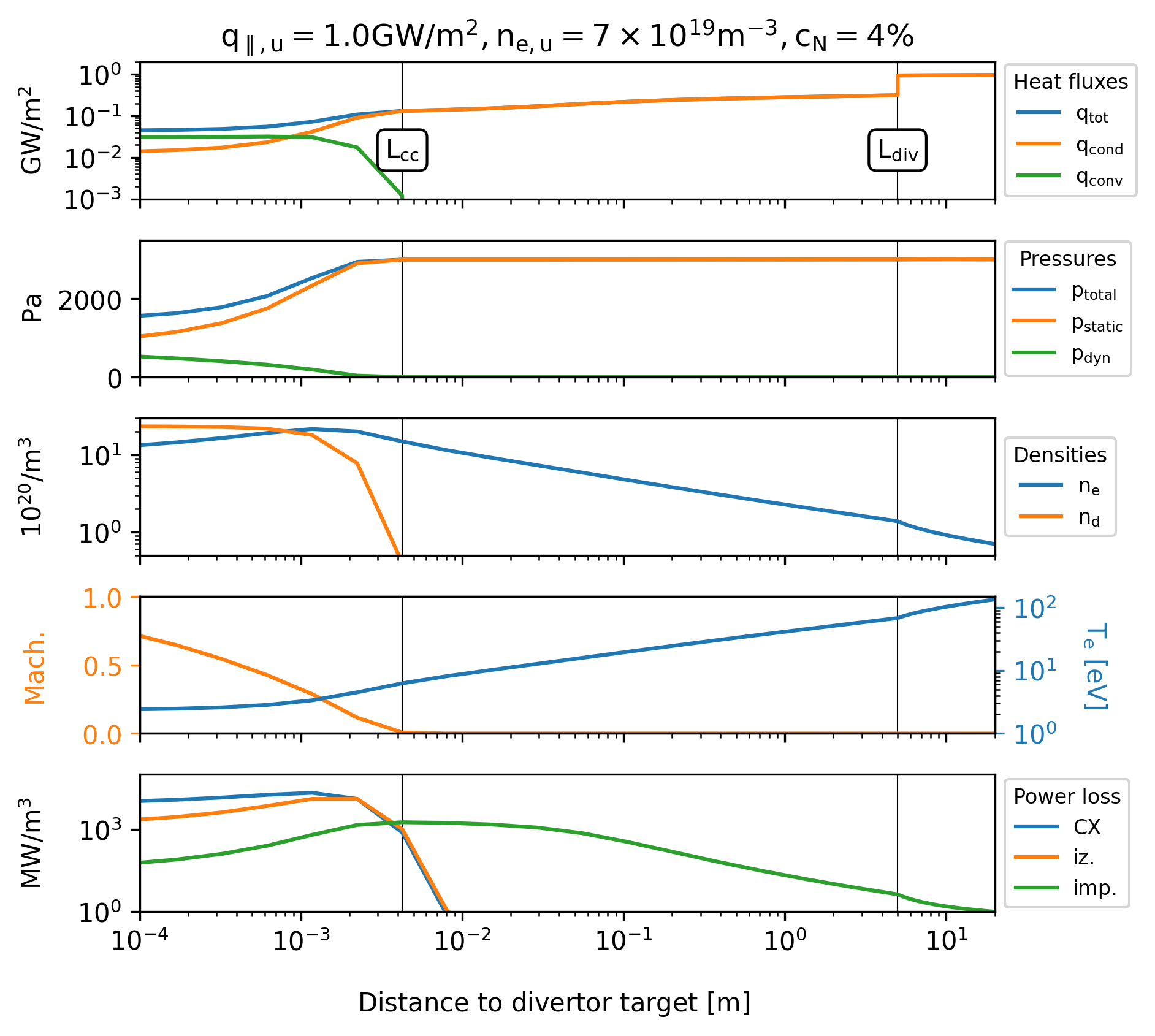}
    \caption{1D profiles computed from the Kallenbach model, for an upstream parallel heat flux of $1\mathrm{GW/m^2}$, an upstream electron density of $7\times10^{19}\mathrm{m^{-3}}$ and a nitrogen concentration of $4\%$. The line marked $L_{cc}$ near $0.4\mathrm{mm}$ indicates the position where $q_{cond}/q_{tot}>0.99$, and the line marked $L_{div}$ at $5.0\mathrm{m}$ indicates the divertor entrance position, where we switch from using $\lambda_{INT}$ to $\lambda_q$. The \textit{top row} gives the total (\textit{blue}), conductive (\textit{orange}) and convective (\textit{green}) heat fluxes. The \textit{second row} gives the total (\textit{blue}), static (\textit{orange}) and dynamic (\textit{green}) pressures. The \textit{third row} gives the electron (\textit{blue}) and neutral-deuterium (\textit{orange}) densities. The \textit{fourth row} gives the electron temperature (\textit{blue, compared to right-side y-axis}) and the Mach number (\textit{orange, compared to the left-side y-axis}). The \textit{bottom row} gives the energy loss due to charge-exchange (\textit{blue}), neutral ionization (\textit{orange}) and impurity radiation (\textit{green}).}
    \label{fig:1D_profile_RCC}
\end{figure}
In figure \ref{fig:1D_profile_RCC}, we show the parallel profiles computed by the Kallenbach model, which we divide into three distinct regions\footnote{The input parameters match those used in figure 4 of Kallenbach et al., 2016 \cite{Kallenbach2016-hk}, to ensure that we recover the same results as the original model. The models are almost identical, although the neutral density drops off slightly faster in the reformulated model since we have disabled the fast neutral flux term.}. Close to the divertor target, the neutral population due to recycling is significant, the total pressure is not conserved, and a significant fraction of the parallel heat flux is carried by heat convection. The parallel profiles vary steeply in this `\textit{convective}' region, up to a point at $x = L_{cc}\sim 0.4\mathrm{mm}$ from the divertor target. We define the end of the convective region as the point where the convective heat flux drops below $1\%$ of the total heat flux, which we refer to as the `\textit{convective-conductive}' boundary. Upstream of the convective-conductive boundary ($x>L_{cc}$), the total pressure is conserved and equal to the static pressure, and the parallel heat flux is carried purely by heat conduction. This `\textit{conductive}' region is further separated into the \textit{divertor} ($x<L_{div}$) and \textit{main chamber} ($x>L_{div}$) at the divertor entrance $x=L_{div}$, where we switch from using the integral heat flux width $\lambda_{INT}$ to using the upstream heat flux width $\lambda_q$. Switching the heat flux width causes a jump in the parallel heat flux density and in the derivatives of the electron temperature and density -- while the profiles themselves are continuous and the pressure is conserved across the divertor entrance.

Since the static pressure is conserved and the heat flux is dominated by electron conduction throughout the conductive region, the assumptions of the Lengyel model should hold throughout the conductive region. However, we can't straightforwardly deal with the discontinuity in the heat flux at the divertor entrance, and as such we first evaluated the Lengyel model in its spatial form, equivalent to using equations \ref{eq:kallenbach_dqtotdx}, \ref{eq:kallenbach_dTedx} and \ref{eq:kallenbach_dndx}, assuming $q_{conv}\to0$ and neglecting neutral ionization, recombination and charge exchange;
 \begin{align}
    \frac{dq_{cond}}{dx} &= n_e^2 c_z L_z(T_e)\label{eq:spatial_leng_dq}\\
    \frac{dT_e}{dx} &= \frac{q_{cond}}{\kappa_{e} T_e^{5/2}}\label{eq:spatial_leng_dTe}\\
    \frac{dn_e}{dx} &= -\frac{n_e}{T_e} \frac{dT_e}{dx}\label{eq:spatial_leng_dn}
\end{align}
As a direct (albeit impractical) proof-of-principle, we integrated the spatial Lengyel equations from the convective-conductive boundary (computed by the Kallenbach model) to the outboard midplane. The upstream parameters from the spatial Lengyel model matched values from the Kallenbach model within $1\%$ for a scan across a broad range of input parameters\footnote{A 3D matrix of 10 log-spaced values of $q_{\perp,t}$ from $0.1$ to $10\mathrm{MW/m^2}$, 10 log-spaced values of $T_{e,t}$ from $2$ to $50\mathrm{eV}$, and 3 $c_z$ values at 1\%, 2\% and 4\%.}, demonstrating that the Lengyel assumptions are indeed valid in the conductive region. Combined with a suitable model for the convective region, we should be able to reproduce the results of the Kallenbach model and scaling in a Lengyel-like model.

\section{Power and momentum loss in the convective region}\label{sec:convective_scalings}

\begin{figure}
    \centering
    \includegraphics[width=0.8\linewidth]{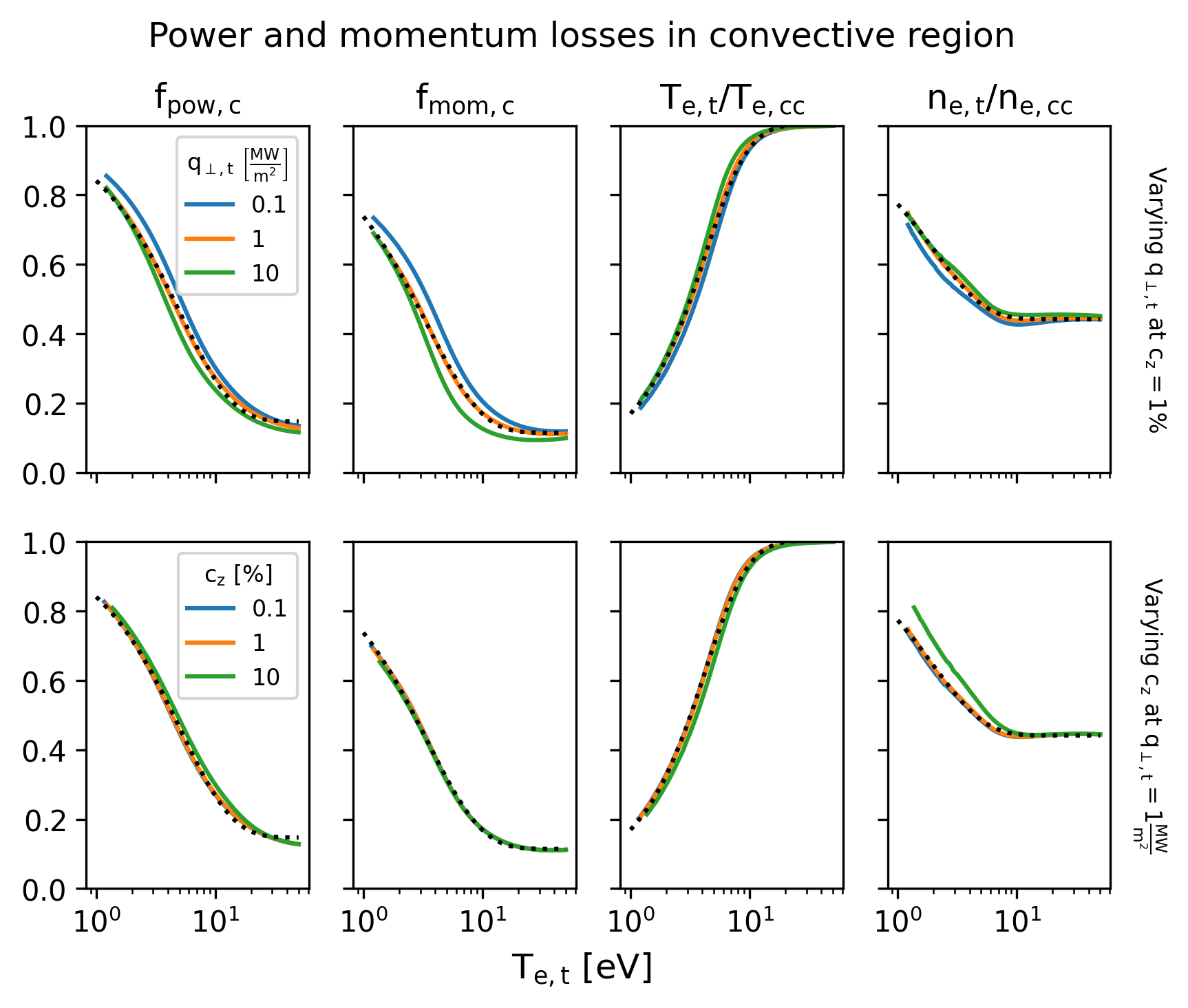}
    \caption{
    Convective region losses and ratios, as functions of the target electron temperature $T_{e,t}$. The \textit{first and second columns} give the power and momentum losses in the convective region, $f_{pow,c}=1-\frac{q_t}{q_{cc}}$ and $f_{mom,c}=1-\frac{2 n_{e,t} T_{e,t}}{n_{e,cc}T_{e,cc}}$ respectively. The \textit{third and fourth columns} give the temperature and density ratios across the convective region. The \textit{top row} gives results for three target perpendicular heat fluxes $q_{\perp,t}$ at $c_z=1\%$. The \textit{bottom row} gives results for three impurity concentrations $c_z$ at $q_{\perp,t}=1\mathrm{MW/m^2}$. For each subplot, a fit is given as a \textit{black, dotted line}, with fit parameters given in table \ref{tab:fit_parameters}.}
    \label{fig:convective_power_and_mom_loss}
\end{figure}
We need a model for the power and momentum loss across the convective region, since the Lengyel model is not valid in that region. We could use the two-region model from Siccinio et al., 2016 \cite{Siccinio2016-cp}. However, this model assumes that the total pressure is constant and that the heat flux is entirely convective in the convective region\footnote{In the Siccinio model, the convective-conductive boundary is defined at a critical temperature $T_C\sim 15\mathrm{eV}$, whereas we use $q_{conv}/q_{tot}=1\%$. The Siccinio definition gives a larger convective region unless $T_{e,t}\geq T_C$, and therefore our definition for the convective region gives a necessary (but not sufficient) test for the validity of the Siccinio assumptions provided that $T_{e,t}<T_C$.}, in contrast to what we observe in figure \ref{fig:1D_profile_RCC}. Rather than develop a reduced analytical model for the convective region losses, we used the Kallenbach model to derive heuristic scalings. We performed scans to investigate how the convective losses change as we vary the target electron temperature $T_{e,t}$, the target heat flux $q_{\perp,t}$ and the impurity fraction $c_z$. For each point in the scan, we calculated the total momentum loss (the ratio of the target and upstream total pressures, equation 19 of Stangeby, 2018 \cite{Stangeby2018-gi})
\begin{align}
    f_{mom} = 1 - \frac{p_{tot,t}}{p_{tot,u}}
\end{align}
We can rewrite this in terms of the density and temperature at the convective-conductive boundary\footnote{Since the total pressure is conserved in the conductive region and the pressure at the convective-conductive boundary is entirely static, $p_{tot,u}=p_{tot,cc}=n_{e,cc}(T_{e,cc} + T_{i,cc})=2n_{e,cc}T_{e,cc}$. At the target, our boundary condition (equation \ref{eq:kallenbach_vt}) forces the ion velocity to the sound speed, and therefore $p_{tot,t}=n_{e,t}(T_{e,t} + T_{i,t}) + m_i n c_{s,t}^2= 4 n_{e,t}T_{e,t}$.
} $n_{e,cc}$ and $T_{e,cc}$ as 
\begin{align}
    f_{mom}&=1 - 2 \frac{n_{e,t}}{n_{e,cc}}\frac{T_{e,t}}{T_{e,cc}}
\end{align}
We also compute the total power loss across the convective region in terms of the total target parallel heat flux density $q_{\parallel,tot,t}$ and the total parallel heat flux density at the convection-conduction interface $q_{\parallel,tot,cc}$
\begin{align}
    f_{pow,c}=1-\frac{q_{\parallel,tot,t}}{q_{\parallel,tot,cc}}\label{eq:fpowc}
\end{align}
The momentum loss $f_{mom}$, power loss $f_{pow,c}$, temperature ratio $T_{e,t}/T_{e,cc}$ and density ratio $n_{e,t}/n_{e,cc}$ across the convective layer are shown in figure \ref{fig:convective_power_and_mom_loss}, as functions of the target electron temperature. Varying the target heat flux and impurity fraction by a factor-of-ten in each direction has a much less significant effect than varying the target electron temperature, and as such we fit curves for a fixed $q_{\perp,t}=1\mathrm{MW/m^2}$ and $c_N=1\%$ using the fit function developed for momentum loss in section 6 of Stangeby, 2018 \cite{Stangeby2018-gi}
\begin{align}
    f_{mom} = 1 - A \left( 1 - \exp\left[ \frac{-T_{e,t}}{w}\right] \right)^s
    \label{eq:tetar_fit_function}
\end{align}
The fits are marked in figure \ref{fig:convective_power_and_mom_loss}, and the fit parameters are given in table \ref{tab:fit_parameters}. For consistency, we calculate the temperature ratio as $T_{e,t}/T_{e,cc}=\frac{1 - f_{mom}}{2 n_{e,t}/n_{e,cc}}$. Using these fits for the power and momentum loss in the convective region, the upstream parameters from the spatial Lengyel model matched within $10\%$ for $90\%$ of the points in a scan across a broad range of input parameters\footnote{The same 3D matrix scan used before, with 10 log-spaced values of $q_{\perp,t}$ from $0.1$ to $10\mathrm{MW/m^2}$, 10 log-spaced values of $T_{e,t}$ from $2$ to $50\mathrm{eV}$, and 3 $c_z$ values at 1\%, 2\% and 4\%.}, and within $20\%$ for all points in the scan. Although not perfect, this level of accuracy is sufficient for our model. The worst agreement is found for points with $q_{\perp,t}$ values either much lower or much higher than the $q_{\perp,t}=1\mathrm{MW/m^2}$ for which we developed the convective loss fits. As such, we recommend developing new convective loss fits\footnote{New convective loss fits can be made by adjusting the input parameters for \texttt{rcc\_target\_electron\_temp\_scan\_for\_fit} in the notebook `Studying the convective layer' (see section \ref{sec:software_availability}).} if performing analyses with the target heat flux outside the range $0.1-10\mathrm{MW/m^2}$.
\begin{table}
    \centering
    \begin{tabular}{cccc}
         &  $f_{pow,c}$&  $f_{mom,c}$& $n_{e,t}/n_{e,cc}$\\
         $A$&  $0.853 \pm 0.002$&  $0.886 \pm 0.0009$& $0.559 \pm 0.0007$\\
         $w$ $[eV]$&  $5.2 \pm 0.08$&  $3.83 \pm 0.04$& $2.02 \pm 0.03$\\
         $s$&  $0.964 \pm 0.01$&  $0.828 \pm 0.007$& $0.96 \pm 0.02$\\
    \end{tabular}
    \caption{Fit parameters for the convective region power loss $f_{pow,c}$, momentum loss $f_{mom,c}$ and density ratio $n_{e,t}/n_{e,cc}$, as shown in figure \ref{fig:convective_power_and_mom_loss}. These fits are given for a $1\%$ nitrogen impurity fraction at a perpendicular target heat flux of $1\mathrm{MW/m^2}$.}
    \label{tab:fit_parameters}
\end{table}
\section{Deriving an extended Lengyel model}\label{sec:extended_lengyel}

Combining the Lengyel equations with fits for the convective-region power and momentum loss, we should be able to reproduce the results of the Kallenbach model. We've already shown that this is possible using the spatial Lengyel model. However, the spatial Lengyel model is defined using coupled differential equations and takes target parameters as inputs, which makes it somewhat unwieldy. In this section, we extend the semi-analytical derivation of Lengyel \& Goedheer, 1981 \cite{Lengyel1981-in} to account for convective region losses and divertor broadening to derive a simpler, more intuitive `extended Lengyel model'. We combine equations \ref{eq:spatial_leng_dq} and \ref{eq:spatial_leng_dTe} to write
\begin{align}
    q\frac{dq}{dx} &= \kappa_{e}T_e^{5/2} \frac{dT_e}{dx}n_e^2 c_z L_z(T_e)
\end{align}
where we write $q=q_{cond}$ since we assume all heat flux is conducted in the conductive region. We can integrate this equation along the field-line, from two arbitrary points $a$ and $b$
\begin{align}
    \int_a^b q\frac{dq}{dx}dx &= \int_a^b \kappa_{e}T_e^{5/2} \frac{dT_e}{dx}n_e^2 c_z L_z(T_e)dx\\
    \implies q_b^2 - q_a^2 &=2\kappa_{e}n_{e,u}^2 T_{e,u}^2c_z L_{INT}^{a \to b} \label{eq:lengyel_semianalytical}
\end{align}
where
\begin{align}
L_{INT}^{a \to b}&=\int_{T_a}^{T_b} L_z(T_e) \sqrt{T_e} dT_e
\end{align}
and where we assume $n_e(x)T_e(x)=n_{e,u}T_{e,u}$ due to static pressure conservation. Equation \ref{eq:lengyel_semianalytical}  is typically evaluated from the target $t$ to the upstream outboard-midplane $u$ and solved for $c_z$, giving the basic Lengyel model (equation \ref{eq:basic_lengyel}). However, since the Lengyel model equations are only valid in the conduction-dominated region, we need to account for momentum and power loss in the convective region. In addition, the heat flux width is broadened in the divertor due to cross-field transport. To account for this, in the Kallenbach model the heat flux width switches from $\lambda_{INT}$ to $\lambda_q$ at the divertor entrance $L_{div}$ which introduces a discontinuity in $\frac{dq}{dx}$. To account for these effects, we evaluate equation \ref{eq:lengyel_semianalytical} twice; once from the conductive-convective $cc$ boundary to the divertor entrance $div$, and then again from the divertor entrance $div$ to the upstream point $u$. We use the broadening factor
\begin{align}
b=\frac{\lambda_{INT}}{\lambda_q}\approx1 + \frac{1.64 S}{\lambda_q}\label{eq:broadening_factor}
\end{align}
from Kallenbach et al., 2018 \cite{Kallenbach2018-ct}. For this work we use a constant value of $b=3$, equivalent to setting $S=1.22\lambda_q$, matching the value used in Kallenbach et al., 2016 \cite{Kallenbach2016-hk}. Using this notation, we write the heat flux at the divertor entrance as
\begin{align}
    \lim_{x\to L_{div}^-} q(x) & =q_{div}/b \text{ if approaching from downstream}\\
    \lim_{x\to L_{div}^+} q(x) & =q_{div} \text{ if approaching from upstream}
\end{align}
We then write two separate Lengyel equations, above and below the X-point,
\begin{align}
    \left(q_{div}/b\right)^2 - q_{cc}^2 &= 2\kappa_{e}n_{e,u}^2 T_{e,u}^2c_z L_{INT}^{cc \to div}\\
    q_u^2 - q_{div}^2 &= 2\kappa_{e}n_{e,u}^2 T_{e,u}^2c_z L_{INT}^{div \to u}
\end{align}
Dividing one equation by the other and solving for $q_{div}^2$ gives
\begin{align}
    q_{div}^2 =(1-f_{rad,main})^2 q_u^2 &= \frac{L_{INT}^{cc \to div} q_{u}^2 + L_{INT}^{div \to u} q_{cc}^2}
    {L_{INT}^{cc \to div} + L_{INT}^{div \to u} / b^2}\label{eq:fradmain}
\end{align}
Adding the two Lengyel equations together, we combine the two temperature integrals to write
\begin{align}
    q_u^2 + \left(\frac{1}{b^2} -1\right) q_{div}^2 - q_{cc}^2 &= 2\kappa_{e}n_{e,u}^2 T_{e,u}^2 c_z L_{INT}^{cc \to u}
\end{align}
which we then solve for the impurity fraction, giving
\begin{align}
    \Aboxed{c_z &= \frac{q_u^2 + \left(\frac{1}{b^2} - 1\right) q_{div}^2 - q_{cc}^2}{2\kappa_{e}n_{e,u}^2 T_{e,u}^2 L_{INT}^{cc \to u}}}\label{eq:extended_lengyel}
\end{align}
or equivalently
\begin{align}
    c_z &= \frac{\left(1 + \left(\frac{1}{b^2} - 1\right) (1-f_{rad,main})^2 - \left(\frac{1-f_{pow,SOL}}{1-f_{pow,c}}\right)^2\right)q_u^2}{2\kappa_{e}n_{e,u}^2 T_{e,u}^2 L_{INT}^{cc \to u}}\label{eq:extended_lengyel_qu}
\end{align}
for $f_{rad,main}$, $f_{pow,SOL}$ and $f_{pow,c}$ defined by equations \ref{eq:fradmain}, \ref{eq:fradsol} and \ref{eq:fpowc} respectively. To evaluate the temperature integrals, we need the temperatures at the divertor entrance and upstream. If we neglect the change in the parallel heat flux due to radiation, we can integrate equation \ref{eq:spatial_leng_dTe} to find 
\begin{align}
    T_{e,div} &= \left( T_{e,cc}^{7/2} + \frac{7}{2}\frac{(q_u/b) L_{div}}{\kappa_{e}} \right)^{2/7}\label{eq:Tediv_SH}\\
    T_{e,u} &= \left( T_{e,div}^{7/2} + \frac{7}{2}\frac{q_u (L_\parallel - L_{div})}{\kappa_{e}} \right)^{2/7}\label{eq:Teu_SH}
\end{align}
To calculate the heat flux at the conductive-convective interface $q_{cc}$, we also need the target heat flux $q_t$. This is calculated from the total power loss $1-f_{pow,SOL}\equiv\frac{1-f_{rad,SOL}}{b}$ using the two-point-model (equation 15 from ref \cite{Stangeby2018-gi})
\begin{align}
    q_t&=(1-f_{pow,SOL})q_u\\&=\sqrt{\frac{T_{e,t}}{8m_i}}\gamma_{sh}(1-f_{mom})n_{e,u}(T_{e,u}+T_{i,u})\label{eq:fradsol}
\end{align}
Once we have $q_t$, we can then calculate $q_{cc}$ using the convective loss function (equation \ref{eq:fpowc}).

\subsection{Solving the extended Lengyel model} \label{subsec:solving_ext_lengyel}
\begin{figure}
    \centering
    \hspace*{-3cm}
    \includegraphics[width=1.5\linewidth]{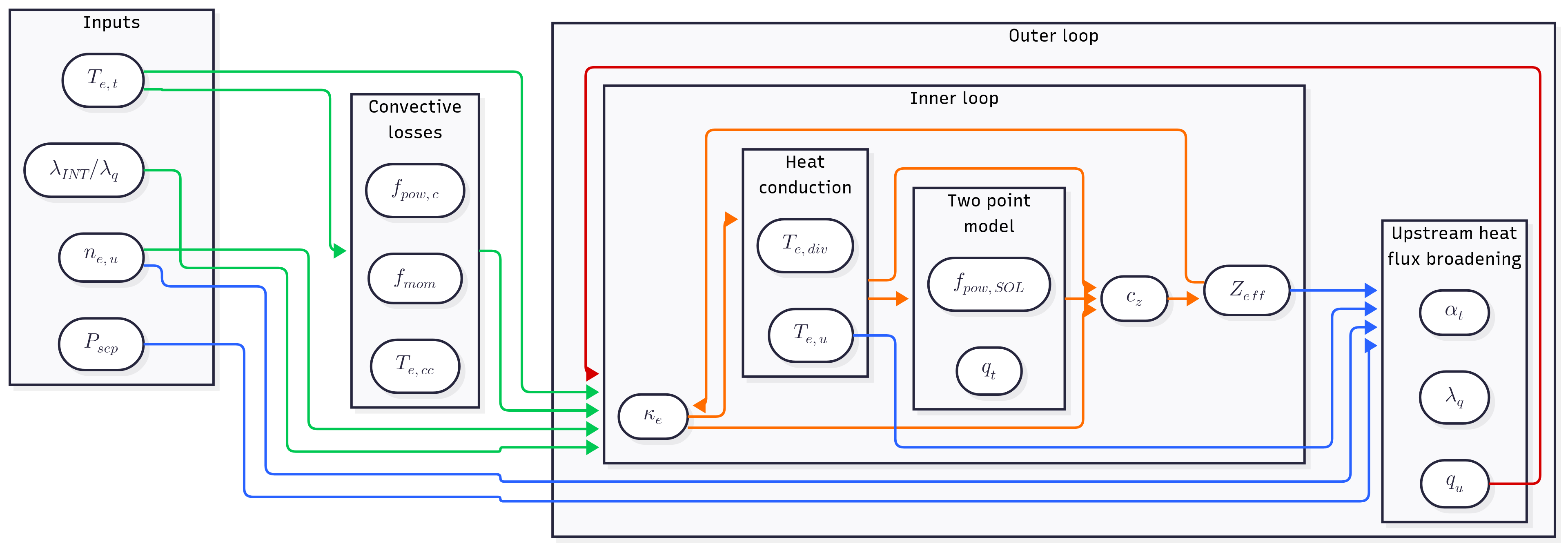}
    \caption{The extended Lengyel model, see discussion in section \ref{subsec:solving_ext_lengyel}.}
    \label{fig:flowchart}
\end{figure}

This gives us a set of equations which we can use to calculate the impurity fraction $c_z$ required to reach a given target electron temperature $T_{e,t}$ or momentum loss fraction $f_{mom}$, for a given divertor broadening $b=\lambda_{INT}/\lambda_q$, upstream electron density $n_{e,u}$ and upstream parallel heat flux density $q_{u}$. In figure \ref{fig:flowchart}, we give an algorithm which solves these equations iteratively. First, we set $T_{e,t}$ or $f_{mom}$ and use the convective fits from section \ref{sec:convective_scalings} to calculate the heat flux and pressure at the $cc$ interface. We then calculate the divertor and upstream temperatures from heat conduction (equations \ref{eq:Tediv_SH}-\ref{eq:Teu_SH}), and use the two point model (equation \ref{eq:fradsol}) to calculate how much power needs to be radiated in the scrape-off-layer. This in turn is passed to the extended Lengyel model (equation \ref{eq:extended_lengyel}) to calculate $c_z$ and $Z_{eff}$. We then update the electron heat conductivity with the new value of $Z_{eff}$ (equation \ref{eq:kappa_z}), and we repeat the \textit{inner loop} until we find a consistent set of $c_z$, $f_{rad,SOL}$ and $T_{e,u}$. The $c_z$ and $T_{e,u}$ values can then be used to calculate the turbulence-broadened heat flux width $\lambda_q$ (using equation \ref{eq:lambda_q_broadening}, which we discuss in section \ref{subsec:calculating_expt_values}) and the upstream heat flux density $q_u$, which is then passed back to the \textit{inner loop} and. The \textit{outer loop} is then iterated until a consistent set of $c_z$, $\alpha_t$, $q_u$ and $T_{e,u}$ is found. We studied the impact of correcting for individual terms by disabling parts of the algorithm above, effectively giving five versions of the Lengyel model. These are are:
\begin{itemize}
    \item the basic Lengyel model with $\lambda_{INT}/\lambda_q = 1$, $\kappa_z=1$, $f_{pow,c}=0$, $T_{e,t}/T_{e,cc}=1$ and $\lambda_q$ fixed, with $f_{mom}(T_{e,t})$ computed from the convective-loss fit,
    \item the $\lambda_{INT}/\lambda_q$-corrected model, which extends the basic Lengyel model by allowing for $\lambda_{INT}/\lambda_q>1$,
    \item the $f_{conv}$-corrected model, which extends the $\lambda_{INT}/\lambda_q$-corrected model by computing $f_{pow,c}$ and $T_{e,t}/T_{e,cc}$ from the convective-loss fits,
    \item the $\kappa_z$-corrected model, which extends the $f_{conv}$-corrected model by computing $\kappa_z$ from $Z_{eff}$ and using the inner loop to find a consistent set of $c_z$, $f_{rad,SOL}$ and $T_{e,u}$,
    \item the $\alpha_t$-corrected model, which extends the $\kappa_z$-corrected model by computing $\lambda_q$ and $q_u$ from $\alpha_t$ (using equation \ref{eq:lambda_q_broadening}) and using the inner and outer loops to find a consistent set of $c_z$, $\alpha_t$, $q_u$ and $T_{e,u}$.
\end{itemize}

\begin{figure}
    \centering
    \includegraphics[width=0.8\linewidth]{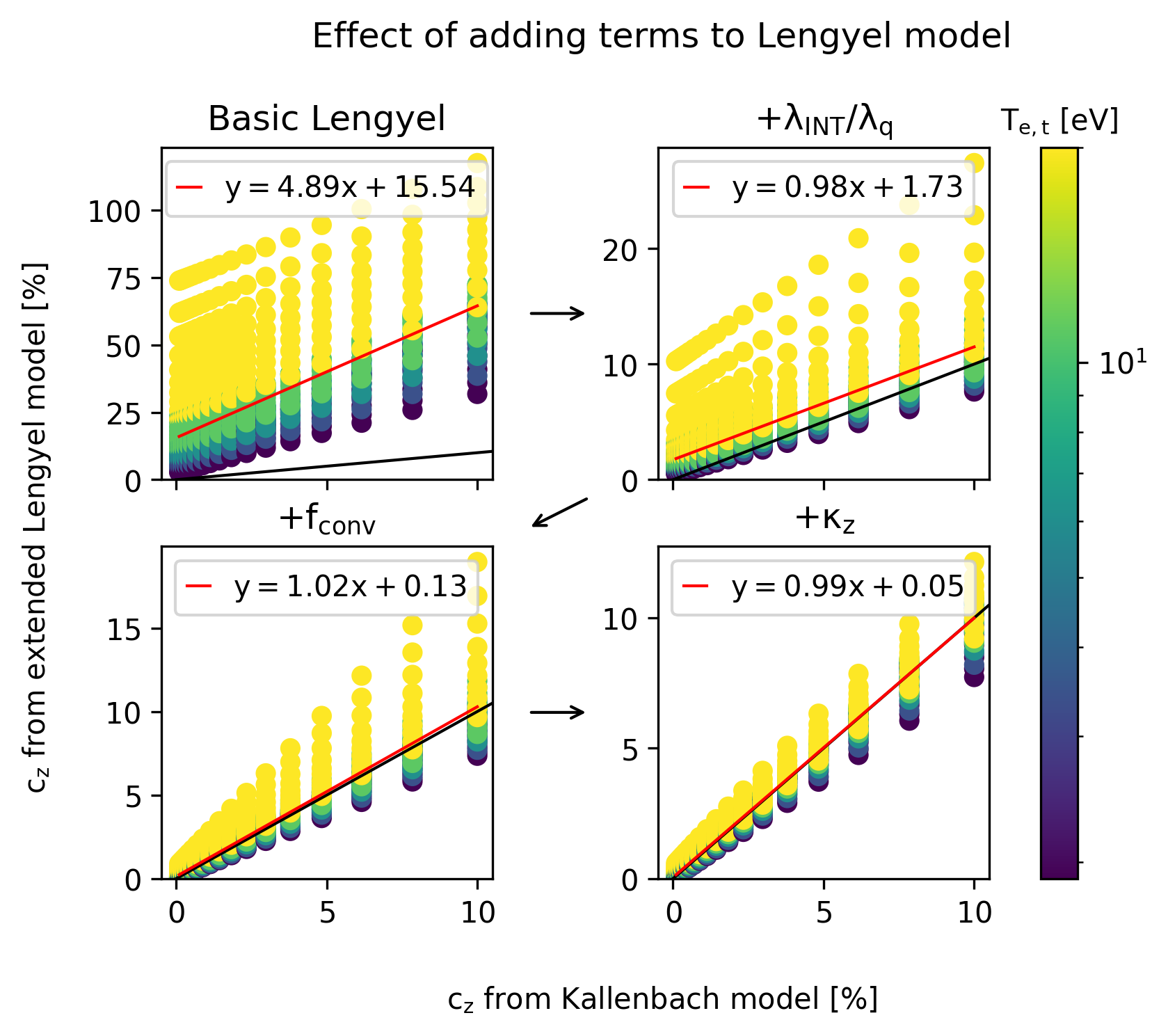}
    \caption{The impurity concentration predicted by the different versions of the extended Lengyel model given in section \ref{subsec:solving_ext_lengyel}. These are compared to the impurity concentration used as input by the Kallenbach model, for a $10\times10\times20$ scan of log-spaced $q_{\perp,t}$, $T_{e,t}$ and $c_z$. The subplots show the basic Lengyel model (\textit{top left}), the $\lambda_{INT}/\lambda_q$-corrected model (\textit{top right}), the $f_{conv}$-corrected model (\textit{bottom left}) and the $\kappa_z$-corrected model (\textit{bottom right}). For each subplot, a $1:1$ line is shown in \textit{black} and a linear regression is shown in \textit{red}, with fit parameters given in the legend. Points are colored according to their target electron temperature.}
    \label{fig:lengyel_comparison}
\end{figure}

In figure \ref{fig:lengyel_comparison}, we show how the different models compare to the Kallenbach model. Starting with the basic Lengyel model, we see that this model predicts much higher impurity concentrations than the Kallenbach model, with both a constant factor of $4.9\times$ and a offset of $+15\%$. We can largely eliminate the factor by correcting for the switch from $\lambda_q$ to $\lambda_{INT}$ at the divertor entrance, which also reduces the offset to $+2\%$. If we also correct for losses in the convective region, this eliminates the remaining offset. Finally, if we iteratively solve for a consistent $c_z$ and $\kappa_z$, we reduce the scatter and further approach a 1:1 match\footnote{We don't compare the $\alpha_t$-corrected model to the Kallenbach model, since the Kallenbach model uses a fixed $\lambda_q$.}. This is a satisfying result: with a small number of corrections, we can reproduce the results of the Kallenbach model with a simple Lengyel-like model.

\section{Comparing to experiment}\label{sec:comparing_to_experiment}

Can we use our extended Lengyel model to interpret experimental results and predict the impurity concentration needed for detachment? In this section, we test our extended model against three experimental results —  the experimentally-determined Kallenbach scaling (equation \ref{eq:kallenbach_scaling}), the density and power dependencies reported in Henderson et al., 2021 \cite{Henderson2021-ts}, and a mixed-impurity experimental data point from Kallenbach et al., 2024 \cite{Kallenbach2024-as}.

\subsection{Calculating experimental values}\label{subsec:calculating_expt_values}
To compare our model to experimental results, we need to convert between the parallel heat flux density $q_\parallel$ and the power crossing the separatrix $P_{sep}$. We assume that the outer divertor receives $f_{odiv}=2/3$ of $P_{sep}$, and focus on the first $\lambda_q$ which receives $1 - 1/e$ of the power to the outer divertor. This power is assumed to be distributed over a ring of width $\lambda_{q,u}$ at the outboard midplane with circumference $2\pi(R+a)$, projected from the poloidal to the parallel direction using the upstream pitch angle $\frac{B_{pol,u}}{B_{tor,u}}$ (the ratio of the poloidal and toroidal field at the outboard midplane). This lets us write
\begin{align}
    q_{\parallel,u} &= \frac{P_{sep}(1 - 1/e)f_{odiv}}{2\pi(R+a)\lambda_{q,u} \frac{B_{pol,u}}{B_{tor,u}}}
    \label{eq:qpar_definition}
\end{align}
for $R$ the major radius and $a$ the minor radius. For the heat flux decay width $\lambda_{q,u}$, we either use a fixed value of $\lambda_{q,u}=1.66\mathrm{mm}$ (matching the value used in Kallenbach et al., 2016 \cite{Kallenbach2016-hk}) or a turbulence-broadened value (equations 2 and 22 from Eich et al., 2020 \cite{Eich2020-jx})
\begin{align}
    \lambda_{q,avg} &= 0.6\rho_{s,pol,avg}\left(1 + 2.1\alpha_t^{1.7} \right)\label{eq:lambda_q_broadening}\\
    \lambda_{q,u} &= \lambda_{q,avg} / \left( \frac{B_{pol,u}}{B_{pol,avg}}\frac{R+a}{R} \right)
\end{align}    
where
\begin{align}
    \frac{B_{pol,u}}{B_{pol,avg}}&\approx 4/3\\
    B_{pol,avg}&=\frac{\mu_0I_p}{2\pi R f_{shaping}}\label{eq:Bpol_avg}\\
    \rho_{s,pol,avg}&= \frac{\sqrt{m_i T_{e,u}}}{e B_{pol,avg}}\\
    \alpha_t&\approx 3.13\times10^{-18}R q_{cyl}^2 \frac{n_{e,u}}{T_{e,u}^2}Z_{eff}\label{eq:alpha_t}\\
    q_{cyl} &= \frac{2\pi a^2 B_{axis}}{\mu_0 I_p R}f_{shaping}^2\\
    f_{shaping} &= \sqrt{\frac{1 + \kappa_{95}^2\left(1 + 2 \delta_{95}^2 - 1.2\delta_{95}^3\right)}{2}}
\end{align}
where subscript-$avg$ indicates quantities which are averaged over the separatrix and subscript-$u$ indicates quantities which are defined upstream (typically the outboard midplane), $\rho_{s,pol}$ is the sound Larmor radius, $q_{cyl}$ is the cylindrical safety factor, $B_{axis}$ is the field on-axis, $I_p$ is the plasma current, and $\kappa_{95}$ and $\delta_{95}$ are the elongation and triangularity at the $\Psi_N=0.95$ flux surface (see \cite{Sauter2016-kh}). This expression reduces to the `heuristic drift model' \cite{Goldston2011-xq,Goldston2015-tr} at low upstream collisionality ($\alpha_t \to 0$) and increases by a factor of $\sim3$ at high upstream collisionality ($\alpha_t\sim1$).

We also need to calculate the divertor neutral pressure from our model. Following the method of Kallenbach et al., 2016 \cite{Kallenbach2016-hk}, we estimate this from the incoming perpendicular target ion flux $\Gamma_{i,\perp,t}=\Gamma_{i,\parallel,t} \sin(\alpha)=n_{e,t}c_{s,t}\sin(\alpha)$, for $\alpha$ the target angle of incidence. We assume that by the time the recycled particle flux reaches the neutral pressure gauge, it has thermalized with the vessel walls and recombined into molecules \cite{Kallenbach2019-ab}. Therefore, we compute the outgoing thermal \textit{molecular} deuterium flux $2\times n_{D_2}\times\frac{1}{4}\sqrt{\frac{8 T_{wall}}{\pi m_{D_2}}}$ (where the factor of two is because the outgoing molecular flux takes two incoming atoms) that balances the incoming ion flux, and associate this with a divertor molecular pressure $p_{div}=n_{D_2}T_{wall}$. This lets us calculate the divertor neutral pressure assuming a wall temperature of $300\mathrm{K}$ and a target angle of incidence of $3^{\circ}$ using:
\begin{align}
    \frac{p_{div}}{n_{D_2}T_{wall}} &\equiv \frac{\Gamma_{i,\parallel,t} \sin(\alpha)}{2\times n_{D_2}\times\frac{1}{4}\sqrt{\frac{8 T_{wall}}{\pi m_{D_2}}}}\\
    p_{div} & \equiv \frac{\Gamma_{i,\parallel,t} \sin(\alpha)}{1.52\times10^{23} \mathrm{m^{-2}s^{-1}/Pa}}
\end{align}
\subsection{Comparing to the Kallenbach scaling}
\begin{figure}
    \centering
    \includegraphics[width=0.8\linewidth]{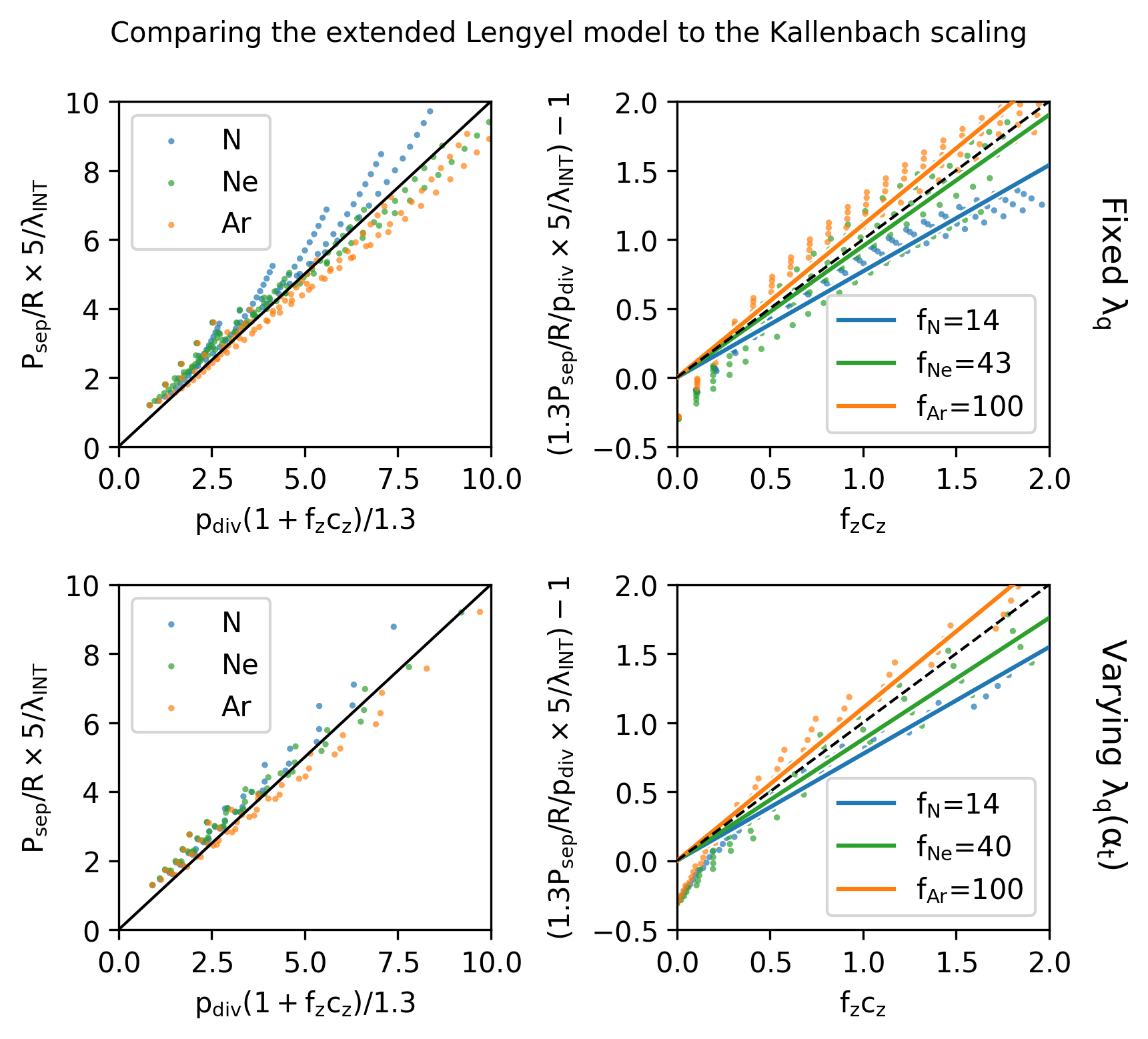}
    \caption{Comparing the extended Lengyel model to the Kallenbach scaling (equation \ref{eq:kallenbach_scaling}), with both a fixed value of $\lambda_q$ (\textit{top row}) and an $\alpha_t$-broadened $\lambda_q$ (\textit{bottom row}), matching the parameters used in figure 7 from Kallenbach et al., 2016 \cite{Kallenbach2016-hk}. The figures compare the left and right hand sides of equations \ref{eq:qdet_1} and \ref{eq:qdet_2}, for nitrogen (\textit{blue}), neon (\textit{green}) and argon (\textit{orange}), for $P_{sep}$ in $\mathrm{MW}$, $R$ in $\mathrm{m}$, $\lambda_{INT}$ in $\mathrm{mm}$, $p_{div}$ in $\mathrm{Pa}$, $c_z$ dimensionless and $f_Z$ defined by equation \ref{eq:fZ_kallenbach}. Perfect agreement with the Kallenbach scaling should give a $1:1$ match indicated by the \textit{solid black line}. The \textit{right subplot legends} give the best-fit $f_Z$ values.}
    \label{fig:comparison_to_kallenbach_scaling}
\end{figure}
To test our model against the Kallenbach scaling, we set $q_{det}=1$ in equation \ref{eq:kallenbach_scaling} (corresponding to partial detachment, or $f_{mom}=0.5$) and rearrange to find
\begin{align}
    \frac{P_{sep} / \mathrm{MW}}{R_0 / \mathrm{m}} \frac{5\mathrm{mm}}{\lambda_{INT}} &= \frac{1}{1.3}\left(1 + f_z c_z\right)\frac{p_{div}}{\mathrm{Pa}}\label{eq:qdet_1}\\
    f_z c_z &= 1.3\frac{P_{sep} / \mathrm{MW}}{R_0 / \mathrm{m}} \frac{5\mathrm{mm}}{\lambda_{INT}}\frac{\mathrm{Pa}}{p_{div}}-1\label{eq:qdet_2}
\end{align}
To see if our model is matching the scaling, we compare the left- and right-hand-sides of these expressions in figure \ref{fig:comparison_to_kallenbach_scaling}, with both fixed $\lambda_q$ (\textit{top row}) and turbulence-broadened (\textit{bottom row}) $\lambda_q$. The models broadly agree with the Kallenbach scaling and with each other. Similar to the comparison to the Kallenbach model (figure 7 from Kallenbach et al., 2016 \cite{Kallenbach2016-hk}), the extended Lengyel models predicts less efficient radiative power dissipation (lower $P_{sep}/(R_0\lambda_{INT}p_{div})$) than the experimental scaling for low values of $c_z$. The best-fit radiative efficiencies $f_N=14$, $f_{Ne}=40-43$, $f_{Ar}=100$ agree fairly closely with the values given in equation \ref{eq:fZ_kallenbach} of $f_N=18,f_{Ne}=45,f_{Ar}=90$. This brings us to our first key result: \textit{the calibration factor between the Lengyel model and experimental scalings is primarily due to heat flux broadening in the divertor, with an additional offset due to momentum and power loss in the recycling region close to the divertor targets}. Accounting for these effects in our extended Lengyel model, we accurately reproduce the empirical Kallenbach detachment onset scaling.

\subsection{Comparing to the Henderson scalings}
\begin{figure}
    \centering
    \includegraphics[width=0.8\linewidth]{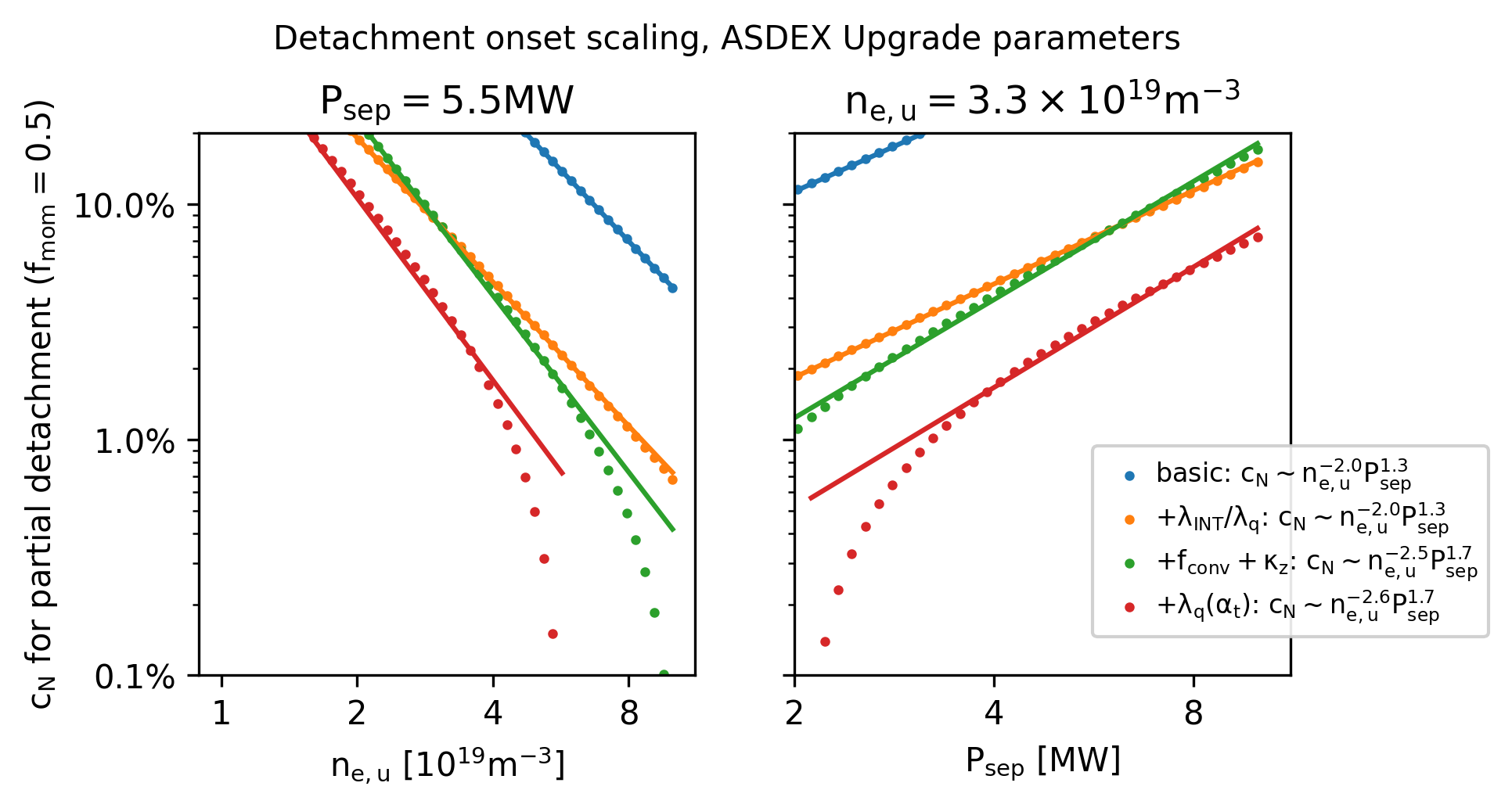}
    \caption{
Nitrogen concentration $c_N$ required to reach partial detachment ($f_{mom}=0.5$) for parameters similar to ASDEX Upgrade with $I_p=0.8\mathrm{MA}$. The \textit{left figure} gives $c_N$ as a function of the upstream density $n_{e,u}$ for $P_{sep}=5.5\mathrm{MW}$, and the \textit{right figure} gives $c_N$ as a function of the power crossing the separatrix $P_{sep}$ for $n_{e,u}=3.3\times10^{19}\mathrm{m^{-3}}$. We show the prediction of several different versions of the Lengyel model -- for the basic model (\textit{blue}), and then adding in corrections for divertor broadening (\textit{orange}), for convective losses and reductions in $\kappa_e$ due to $Z_{eff}$ (\textit{green}), and for $\alpha_t$-broadening (\textit{red}). A power law is fitted for $c_z > 1\%$ for each set of points, with the best-fit given in the legend.}
    \label{fig:alpha_t_broadening}
\end{figure}
In section \ref{sec:introduction}, we estimated that the impurity fraction needed for detachment should scale as $c_z\propto n_{e,u}^{-3.2}$ for the Kallenbach scaling and $c_z \sim {n_{e,u}^{-2}}$  for the Lengyel model. Now that we can reproduce the Kallenbach scaling with a Lengyel-like model, we can use our extended model to see which effects introduce the additional density dependence. In addition to comparing to the density dependence, we also compare to the detachment access scalings reported in Henderson et al., 2023 \cite{Henderson2023-hu}. 

In figure \ref{fig:alpha_t_broadening}, we show the nitrogen concentration required to reach partial detachment on ASDEX Upgrade at $I_p=0.8\mathrm{MA}$, matching the experimental conditions of the largest dataset in Henderson et al., 2023 \cite{Henderson2023-hu}. We see that the basic Lengyel model reproduces the expected $n_{e,u}^{-2}$ scaling. Accounting for divertor broadening keeps the same $n_{e,u}^{-2}$ scaling, but the absolute predicted $c_N$ drops by about a factor of $\sim5$. Further accounting for losses in the convective region\footnote{We also include $Z_{eff}$ corrections for $\kappa_e$ at this point, although these didn't have a strong effect for these parameters so we don't show these separately.}, we find that the results no longer follow a power law. If we restrict our fit to the $c_N>1\%$ region which approximately follows a power-law, we find $n_{e,u}^{-2.5}$. Further allowing $\lambda_q$ to vary with $\alpha_t$, the power-law fit follows $n_{e,u}^{-2.6}$. This is remarkably close to the $c_N\propto n_{e,sep}^{-2.71\pm0.41}$ reported in Henderson et al., 2023 \cite{Henderson2023-hu}. Alternatively, if we perform a best-fit including points that no longer follow a power-law, we can recover the Kallenbach scaling $c_z\propto n_{e,u}^{-3.2}$ if we include points with $0.5\%<c_N<10\%$. This leads to a surprising conclusion; that the impurity concentration required for detachment does not follow a simple power law, and the apparent power law sensitively depends on the exact points used to build a scaling.

We also predict how the nitrogen concentration will vary as a function of the power crossing the separatrix, for a fixed upstream density. Here, the basic Lengyel model matches the Henderson scaling of $c_N\propto (f_{odiv}P_{sep})^{1.24\pm0.45}$, while the extended model predicts a much stronger $c_N\propto (f_{odiv}P_{sep})^{1.7}$ for the $f_{conv}$- and $\alpha_t$-corrected models. The most likely explanation for this disagreement is that the Lengyel model is missing physics which leads to the correct scaling (such as any model for $f_{odiv}$ or $\lambda_{INT}/\lambda_q$), and that the agreement found with the basic Lengyel model is simply fortuitous. However, the good agreement between the basic Lengyel model and the experimental scaling suggests that part of the disagreement could be due to the model used to calculate $T_{e,u}$ in the experimental analysis, which is used to determine $n_{e,u}$ from the edge Thomson system \cite{Kallenbach2016-hk}\footnote{The Henderson scaling uses a relationship between $\lambda_{T_e}$ and $T_{e,u}$ developed in Sun et al., 2017 \cite{Sun2017-ll}, which should already account for upstream broadening, but this approach may not be consistent with equation \ref{eq:lambda_q_broadening}.}. This brings us to our second key result: \textit{the impurity concentration required for detachment onset does not follow simple power laws in terms of the separatrix density or power. By changing the points and parameters used to derive a scaling, we can vary the apparent power law to match the range of existing experimental scalings.}

\subsection{Comparing to experimental data}\label{subsec:comparing_to_experimental_data}

Finally, we demonstrate that our model can be used to quantitatively compute the impurity density needed for detachment onset. For this, we compare to a well-characterized experimental point, ASDEX Upgrade shot \#39520, which is shown in figure 7 of Kallenbach et al., 2024 \cite{Kallenbach2024-as}. We selected a point at $t=5s$, which is partially-detached via combined neon and argon seeding in a 20:1 ratio. At this time-point the plasma current was $I_p=1\mathrm{MA}$, the magnetic field on axis was $B_{axis}=-2.5\mathrm{T}$ and the power crossing the separatrix was $P_{sep}=\mathrm{5.5MW}$, of which $P_{odiv}=2/3P_{sep}=3.66\mathrm{MW}$ was estimated to be flowing towards the outer divertor. The upstream density was estimated to be $n_{e,u}=3.3\times10^{19}\mathrm{m^{-3}}$ from the divertor neutral pressure of $p_{div}=1.9\mathrm{Pa}$ using $n_{e,u}=2.65\times10^{19}\mathrm{m^{-3}}(p_{div}/\mathrm{Pa})^{0.31}$ \cite{Kallenbach2018-ct}, which was consistent with the edge Thomson measurement assuming $T_{e,u}=100\mathrm{eV}$. The target electron temperature was estimated to be $T_{e,t}\approx 2\mathrm{eV}$  from the divertor shunt current\footnote{The target electron temperature is estimated from the total current flowing between the inner and outer divertors (measured by shunt resistors connected to a poloidal row of tiles) \cite{Kallenbach2010-dt}. This current is assumed to be driven by the thermoelectric current, which is related to the difference in sheath-entrance temperatures between the inner and outer divertor \cite{Kallenbach2001-uy}. In cases where the inner divertor detaches (and can therefore be assumed to have a small sheath-entrance temperature) before the outer divertor, the thermoelectric current can be used to calculate the outer divertor sheath-entrance temperature $T_{e,t}$. We use an asymmetric median filter to remove peaks due to ELMs, giving the inter-ELM current to the outer divertor. The filtered current was compared to divertor Langmuir probe measurements to find a calibration factor of $T_{e,t} = 0.02 \mathrm{eV/A}\times I_{therm}$. Using this factor, we estimate an absolute $T_{e,t}$ value from the filtered shunt current measurement.}, corresponding to a momentum loss of $f_{mom}=0.58$ using the fit parameters from table \ref{tab:fit_parameters}. Taking ASDEX Upgrade geometry ($R=1.65\mathrm{m}$, $a=0.5\mathrm{m}$, $\kappa_{95}=1.6$, $\delta_{95}=0.3$, $L_\parallel=20\mathrm{m}$, $L_{div}=5\mathrm{m}$), we calculated a cylindrical safety factor of $q_{cyl}=3.72$, an upstream field-line pitch of $B_{tor,u}/B_{pol,u}=5.15$ and a separatrix-average poloidal field of $B_{pol,avg}=0.29\mathrm{T}$ (giving a separatrix-average poloidal sound Larmor radius of $\rho_{s,pol,avg}=5\mathrm{mm}$ at $100\mathrm{eV}$).

\begin{figure}
    \centering
    \includegraphics[width=0.8\linewidth]{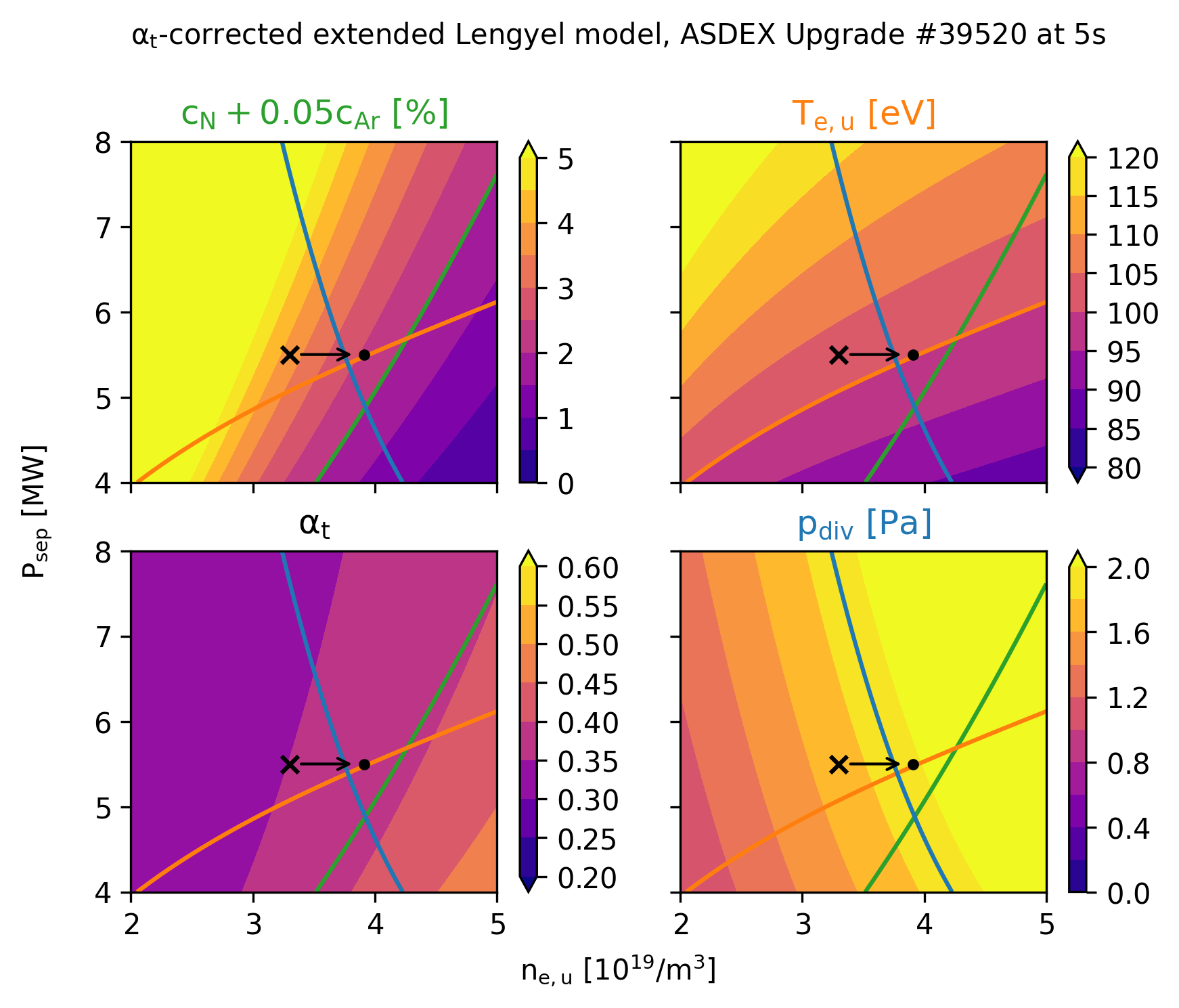}
    \caption{Solutions to the $\alpha_t$-corrected extended Lengyel model, as functions of the upstream electron density $n_{e,u}$ and the power crossing the separatrix $P_{sep}$, for the parameters given in section \ref{subsec:comparing_to_experimental_data}, showing the impurity concentration (\textit{upper left}), upstream temperature (\textit{upper right}), $\alpha_t$ (\textit{lower left}) and divertor neutral pressure (\textit{lower right}). The experimental $n_{e,u}=3.3\times10^{19}\mathrm{m^{-3}}$, $ P_{sep}=\mathrm{5.5MW}$ point is indicated by a \textit{black `x'}, and the experimental impurity concentration, assumed upstream temperature and measured divertor neutral pressure are indicated by \textit{green}, \textit{orange} and \textit{blue lines} respectively.  The \textit{black dot} has $n_{e,u}=3.9\times10^{19}\mathrm{m^{-3}}$, $ P_{sep}=\mathrm{5.5MW}$, selected such that $T_{e,u}=100\mathrm{eV}$.}
    \label{fig:NF2024}
\end{figure}

We assumed a divertor broadening factor of $\lambda_{INT}/\lambda_q=3$, completing the set of inputs needed to run the $\alpha_t$-corrected model discussed in section \ref{subsec:solving_ext_lengyel}. This model predicts that we should reach $T_{e,t}=2\mathrm{eV}$ with a nitrogen concentration of $c_{N}=3.8\%$ and an argon concentration of $c_{Ar}=0.19\%$. This is within a factor of two of the impurity concentration measured by ion spectroscopy, which finds $c_N=2\pm0.5\%$ and $c_{Ar}=0.1\pm0.05\%$. The upstream temperature is predicted to be $T_{e,u}=103\mathrm{eV}$ (leading to $\alpha_t=0.35$ and an upstream heat flux width of $\lambda_{q,u}=2.4\mathrm{mm}$) and the divertor neutral pressure is predicted to be $1.7\mathrm{Pa}$. These values are remarkably close to the assumed upstream temperature and measured divertor pressure from the experiment. As shown in figure \ref{fig:NF2024}, we can improve the match to experiment by increasing the upstream density to $n_{e,u}=3.9\times10^{19}\mathrm{m^{-3}}$ to match the assumed $T_{e,u}=100\mathrm{eV}$. At this point, the impurity concentration, upstream density and divertor neutral pressure all match within uncertainty. Therefore, as well as matching the Kallenbach model and observed detachment onset scalings, this initial result suggests that \textit{the extended Lengyel model quantitatively matches the impurity concentration, separatrix conditions and divertor neutral pressure required for detachment onset}.

\section{Conclusion}\label{sec:conclusion}

The basic Lengyel model for detachment access is simple, easy to implement and intuitive. However, this model predicts an approximately $\sim5\times$ higher impurity concentration than what is actually needed to reach detachment in ASDEX Upgrade experiments \cite{Henderson2021-ts} or in SOLPS simulations \cite{Moulton2021-id}. The Kallenbach scaling provides a far more accurate estimate on several existing devices \cite{Henderson2021-ts}, but since it is semi-empirical it is unclear if this scaling can be extrapolated from existing devices to next step devices such as SPARC and ITER. In this work, we extended the Lengyel model to match and explain the Kallenbach scaling by accounting for broadening of the heat flux width in the divertor, and for power and momentum losses due to neutral ionization near the divertor targets. Further accounting for turbulence-driven broadening of the upstream $\lambda_q$, we reproduced experimentally-determined detachment onset scalings and found a good quantitative match to a given experimental data point.
These results resolve two key puzzles for detachment onset. First, we have shown that the constant-factor disagreement found when comparing Lengyel-like models to experiment \cite{Henderson2021-ts} or transport modeling \cite{Moulton2021-id,Jarvinen2023-xh} can be resolved via simple corrections for divertor cross-field transport and ionization of recycled neutrals. Second, we have shown that the impurity concentration required for detachment decreases faster than $n_{e,u}^2$ due to these same effects, as well as due to turbulence-driven broadening of $\lambda_{q,u}$. We found that the relationship between density and impurity concentration at detachment onset cannot be described by simple power laws, and as such direct extrapolation of fitted power laws should not be used for predicting detachment access on future devices.\\

To accurately predict detachment access on future devices, we need models which have been extensively validated on existing devices. The initial comparison in this paper suggests that the extended Lengyel model already describes experimental results with impressive accuracy, and these results will be confirmed (or contradicted) by further validation against experiment. Since the results of the extended Lengyel model do not follow simple power-laws, the model should be directly compared to raw experimental data similar to the analysis performed in Henderson et al., 2021 \cite{Henderson2021-ts}. For experiments with edge Thomson data available, an interesting extension would be to use the measured $\lambda_{T,u}$ directly in the extended Lengyel model in place of equation \ref{eq:lambda_q_broadening}, and then using the computed $T_{e,u}$ value to evaluate $n_{e,u}$. This validation process will drive further extensions of the model, such as using high-fidelity transport modeling to improve our estimates for convective-region losses (i.e. section 6 of ref \cite{Stangeby2018-gi}), allowing the divertor and upstream impurity concentrations to vary independently \cite{Siccinio2016-cp}, allowing for smooth variations of $\lambda_{INT}$ along the flux-tube, accounting for flux expansion, introducing predictive models for divertor broadening \cite{Brida2025-cc,Henderson2025-mb} and investigating upstream-broadening under highly-dissipative conditions \cite{Eich2020-jx}. This validation-driven development will improve the predictive accuracy of the extended Lengyel model, helping to develop a simple, accurate model for predictive core-edge integration in next-step tokamaks.

\section{Software availability}\label{sec:software_availability}

The software developed for this paper is available at\\\href{https://github.com/cfs-energy/extended-lengyel}{\texttt{https://github.com/cfs-energy/extended-lengyel}}.
The analysis performed for the initial submission of this article is tagged as \texttt{initial\_release}, subsequent revisions are tagged as \texttt{revision\_\#} and the version corresponding to the final article is tagged as \texttt{v1}.

\printbibliography

\end{document}